# Precise physical parameters of three late-type eclipsing binary giant stars in the Large Magellanic Cloud

G. Rojas García[1], D. Graczyk[2], G. Pietrzyński[1], C. Gałan[1], W. Gieren[3], I. Thompson[4], K. Suchomska[1], M. Kałuszyński[1], I. Soszyński[5], A. Udalski[5], P. Karczmarek[3], W. Narloch[1], M. Górski[1], P. Wielgórski[1], B. Zgirski[3], N. Miller[6], G. Hajdu[1], B. Pilecki[1], M. Taormina[1], and M. Lewis[7]

[1] Nicolaus Copernicus Astronomical Centre, Bartycka 18, 00-716 Warsaw, Poland; e-mail: `rojas@camk.edu.pl`
[2] Nicolaus Copernicus Astronomical Centre, Rabiańska 8, 87-100 Toruń, Poland
[3] Departamento de Astronomía, Universidad de Concepción, Casilla 160-C, Concepción, Chile
[4] Carnegie Observatories, 813 Santa Barbara Street, Pasadena, CA 91101-1292, USA
[5] Astronomical Observatory, University of Warsaw, Al. Ujazdowskie 4, 00-478 Warsaw, Poland
[6] Department of Physics and Astronomy, Uppsala University, Box 516, SE-751 20 Uppsala, Sweden
[7] Leiden Observatory, Leiden University, P.O. Box 9513, 2300 RA Leiden, The Netherlands

October 24, 2024

**ABSTRACT**

*Context.* Detached eclipsing binaries (DEBs) allow for the possibility of precise characterization of its stellar components. They offer a unique opportunity to derive their physical parameters in a near-model-independent way for a number of systems consisting of late-type giant stars. Here we aim to expand the sample of low-metallicity late-type giant stars with precise parameters determined.
*Aims.* We aim to determine the fundamental parameters like the mass, radius, or effective temperature for three long-period late-type eclipsing binaries from the Large Magellanic Cloud: OGLE-LMC-ECL-25304, OGLE-LMC-ECL-28283, and OGLE-IV LMC554.19.81. Subsequently we aim to determine the evolutionary stages of the systems.
*Methods.* We fit the light curves from the OGLE project and radial velocity curves from high resolution spectrographs using the Wilson-Devinney code. The spectral analysis was performed with the GSSP code and resulted in the determination of atmospheric parameters such as effective temperatures and metallicities. We used isochrones provided by the MIST models based on the MESA code to derive evolutionary status of the stars.
*Results.* We present the first analysis of three DEBs composed of similar helium burning late-type stars passing through the blue loop. Estimated masses for OGLE-LMC-ECL-29293 (G4III + G4III) are $M_1 = 2.898 \pm 0.031$ and $M_2 = 3.153 \pm 0.038$ $M_\odot$, stellar radii are $R_1 = 19.43 \pm 0.31$ and $R_2 = 19.30 \pm 0.31$ $R_\odot$. OGLE-LMC-ECL-25304 (G4III + G5III) has stellar masses of $M_1 = 3.267 \pm 0.028$ and $M_2 = 3.229 \pm 0.029$ $M_\odot$, radii of $R_1 = 23.62 \pm 0.42$ and $R_2 = 25.10 \pm 0.43$ $R_\odot$. OGLE-IV LMC554.19.81 (G2III + G2III) have masses of $M_1 = 3.165 \pm 0.020$ and $M_2 = 3.184 \pm 0.020$ $M_\odot$, radii of $R_1 = 18.86 \pm 0.26$ and $R_2 = 19.64 \pm 0.26$ $R_\odot$. All masses were determined with a precision better than 2% and radii better than 1.5%. The ages of the stars are in the range of 270-341 Myr.

**Key words.** binaries: eclipsing, spectroscopic – stars: fundamental parameters, orbital parameters – stars: individual: OGLE-LMC-ECL-25304, OGLE-LMC-ECL-29293, LMC554.19.81

## 1. Introduction

A most remarkable feature of double-lined detached eclipsing binaries (DEB) is the possibility to determine their stellar and orbital parameters with high precision solely from observations and empirical calibrations without many model assumptions. For this reason, DEBs serve as an important tool in a wide range of topics in stellar astrophysics, especially as a fundamental source of measurements of stellar radii and masses (e.g. Torres et al. 2010; Southworth 2015; Eker et al. 2018). They are very useful for testing stellar evolution codes (e.g. Higl & Weiss 2017; del Burgo & Allende Prieto 2018), and also for calibrating relations in asteroseismology (Gaulme et al. 2016) as well as surface-brightness relations (Graczyk et al. 2021), among others. During the last two decades, DEBs continued to rise in prominence when it was shown that they allowed for near-geometrical, one-step distance measurements to the Milky Way globular clusters (e.g. Thompson et al. 2001; Rozyczka et al. 2022) and the Local Group galaxies (e.g. Bonanos et al. 2006; Vilardell et al. 2010; Pietrzyński et al. 2013; Graczyk et al. 2014). This use of DEBs was already envisioned by Paczynski (1997) who proposed that DEBs could be well-suited to measure model-independent extragalactic distances to within 1% precision. This ambitious goal has been achieved for the Large Magellanic Cloud (LMC, (Pietrzyński et al. 2019)) using a sample of unique DEBs consisting of two late-type giant stars.

In total 20 late-type giant DEBs were analysed and characterized in the LMC (Graczyk et al. 2018) and 15 such systems in the Small Magellanic Cloud (SMC, Graczyk et al. (2020)). The components of those DEBs are a mixture of red-giant branch stars, horizontal branch stars, and hot asymptotic branch stars. Such binary systems are rare because 1) their initial separation must be large enough to accommodate components within their Roche lobes due to swelling on the red-giant branch and 2) the mass ratio of components must be close to unity in order to observe two giant stars at the same time. The systems from the LMC were utilized not only for distance determination but also to determine the geometry of the LMC disc (Pietrzyński et al. 2019), the age-metallicity relation in the LMC (Graczyk et al.

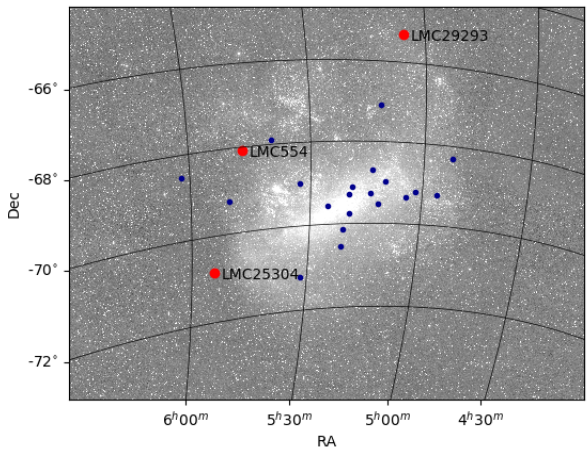

**Fig. 1.** The location of our targets (red dots) on the projection of the Large Magellanic Cloud. Blue dots correspond to the systems studied by Graczyk et al. (2018). The underlying image was obtained by The All Sky Automated Survey and originates from Udalski et al. (2008).

2018) and to constrain the convective core-overshooting in intermediate mass stars (Claret & Torres 2019). Using published OGLE catalogs of eclipsing binary stars in the LMC (Graczyk et al. 2011; Pawlak et al. 2016) we commence a search for additional late-type giant DEBs in the LMC. The main aim is to expand the sample of DEBs suitable to better constrain the disc geometry and eventually to improve the precision of the distance determination to the LMC. The importance of this galaxy is that it is one of only three geometric anchors of the extragalactic distance scale (e.g. Riess et al. 2022).

Here, we present a determination of fundamental physical parameters for three new DEBs consisting of late-type giant stars. All are members of the LMC: OGLE-LMC-ECL-29293, OGLE-LMC-ECL-25304 and OGLE-IV LMC554.19.81 (hereafter LMC29293, LMC25304 and LMC554, respectively). All three eclipsing binaries are well detached and show deep, partial eclipses. Their orbits are circular or only slightly eccentric.

The structure of this work is the following: in Section 2 we present the observational data source we used to model our systems. Section 3 contains information about the radial velocity determination from the spectroscopic data. In Section 4 we describe how we handle the reddening to the sources and how we estimated their preliminary temperatures. Section 5 describes our strategy for a solution to the radial velocity and light curves of the systems. The determination of atmospheric parameters from the disentangled spectra of the stars is addressed in Section 6. The adopted solutions for each DEBs are described in Section 7. Section 8 describes the determination of the evolutionary stage and age of the systems. Section 9 contains our conclusions.

## 2. Observations and data

Basic information about the targets is summarized in Table 1 and their location on the sky is shown in Figure 1. The three systems under study (red circles) were selected because of their long orbital periods and late spectral types. They are located at a relatively large projected distance from the LMC center. The systems were identified through the third and fourth phase of the Optical Gravitational Lensing Experiment (OGLE) (Udalski et al. 1997; Udalski 2003; Udalski et al. 2015). The OGLE project has been conducted for more than two decades with the 1.3m Warsaw Telescope located at Las Campanas Observatory, Chile. For the analysis performed in this article, we use the published I-band light curves. For one of the objects, LMC554, we present an unpublished I-band light curve from the OGLE-IV phase (Soszyński, private communication). The typical cadence of the I-band light curves is between 1 and 3 days. V-band light curves are much more sparse and the light curves in this band are poorly populated, therefore we decided not to include the V-band light curve for the analysis.

High resolution spectroscopic observations of the targets were obtained using three different spectrographs: The Magellan Inamori Kyocera Echelle (MIKE), the Ultraviolet and Visual Echelle Spectrograph (UVES), and the High Accuracy Radial velocity Planet Searcher (HARPS). MIKE is a high resolution double echelle spectrograph implemented at the Magellan Clay 6.5m Telescope at Las Campanas Observatory in Chile; a description of this instrument can be found in Bernstein et al. (2003). With this instrument, it is possible to reach a resolution around $R \sim 40,000$. HARPS (Mayor et al. (2003)) is another high resolution spectrograph mounted on the 3.6m telescope at La Silla Observatory in Chile. The observations were performed in the EGGS mode, reaching a resolution about 80,000. The UVES spectrograph is one of the instruments hosted at the 8.4m UT2 Kueyen telescope being a part of the Very Large Telescope (VLT), at Cerro Paranal Observatory, Chile. This instrument can also reach a resolution of about 80,000 in its high resolution mode. Information on this specific instrument can be found in Dekker et al. (2000).

## 3. Radial velocity determination

The spectra collected with the aforementioned instruments were used to determine radial velocity changes for each source. To achieve this goal we used the broadening function technique (BF), a formalism introduced by Rucinski (1999). This technique is implemented in the RaveSpan software developed by Pilecki et al. (2013). The determination of the RVs using this approach requires templates from the synthetic stellar spectral library of Coelho et al. (2005). The selection of the templates needs to resemble the atmospheric characteristic of the observed sources in terms of metallicity, $\log g$, and the effective temperature. As a value for the metallicity, we used an average metallicity of the LMC as [Fe/H]=-0.42 dex, as reported by Choudhury et al. (2021). Temperatures were set according to a method described in Section 4.

The radial velocities are corrected by an instrumental shift term to account for the use of the three different instruments. This is crucial to ensure consistency by addressing calibration differences, such as those in optical or instrumental setups, allowing accurate comparison of radial velocities. We established the systemic velocity as the one determined from the HARPS spectra, then we applied shifts to the RV determined from UVES and MIKE spectra. For LMC29293 we shift the RV determined with MIKE by 2.122 km/s, and to the UVES by -2.261 km/s. In the case of LMC25304 we apply a shift of 1.187 km/s to the RV determined with MIKE and 0.299 km/s for those we got from UVES spectra. Finally, in the case of LMC554.19.81 the shift applied to the MIKE radial velocities was 0.5 km/s and for the UVES ones 1.140 km/s. The radial velocities corrected by the instrumental shifts are shown in Tables 2, 3, and 4. In the case of MIKE spectra, we used only the ones collected with the red arm of the spectrograph (MIKE-Red). We decided to use this region

**Table 1.** Preliminary information of the targets

|  | **LMC29293** | Ref. | **LMC25304** | Ref. | **LMC554** | Ref. |
|---|---|---|---|---|---|---|
| R.A. (J2000) | 05:05:37.37 | 1 | 05:54:19.44 | 1 | 05:47:08.27 | 4 |
| Decl. (J2000) | −65:17:13.4 | 1 | −70:50:24.0 | 1 | −68:04:12.0 | 4 |
| Period (days) | 123.17 | 3 | 194.517 | 3 | 133.17 | 4 |
| $T_0$ (HJD-2450000) | 5573.83022 | 3 | 7146.40980 | 3 | 5565.68789 | 4 |
| $B$ (mag) | 17.818±0.036 | 2 | 17.461±0.034 | 2 | 17.708±0.035 | 2 |
| $V$ (mag) | 16.904±0.019 | 3 | 16.528±0.021 | 3 | 16.771±0.043 | 2 |
| $I$ (mag) | 15.971±0.020 | 3 | 15.527±0.018 | 3 | 16.039±0.035 | 2 |
| $J$ (mag) | 15.36 ± 0.03 | 5 | 14.89 ± 0.02 | 5,6 | 15.27±0.02 | 6 |
| $H$ (mag) | 14.89 ± 0.03 | 5 | 14.36 ± 0.02 | 5,6 | 14.87±0.02 | 6 |
| $K$ (mag) | 14.82 ± 0.04 | 5 | 14.30 ± 0.02 | 5,6 | 14.79±0.03 | 6 |

**Notes.** Sources: 1-SIMBAD database, 2-Zaritsky et al. (2004), 3-OGLE Collection of Variable Stars (Pawlak et al. (2016)), 4-This paper, 5-Infrared magnitudes from 2MASS 6X Point Source Working Database (Cutri et al. (2012)), 6-Infrared magnitudes from Magellanic Clouds Point Source Catalog (Kato et al. (2007)) converted into 2MASS system

because the SNR is significantly better than for the blue part of spectra.

With RaveSpan we also solved the radial velocity curves producing estimates of the eccentricity, radial velocity semi-amplitudes, arguments of periastron, and systemic velocities. These estimation were used as input parameters for later analysis.

We determined spectroscopic light ratios in *V*-band for all systems using the strength of the broadening function peaks. The resulting light ratios were subsequently corrected for the small surface temperature differences of components. The resulting spectroscopic light ratios are 1.04 ± 0.05, 0.99 ± 0.05 and 1.05 ± 0.04 for LMC29293, LMC25304 and LMC554, respectively. This information subsequently turned out to be useful to better constrain light curve solutions because of the correlation between photometric light ratios and radii ratios – see corner plots in Appendix A.

**Table 2.** Radial velocity measurements for the object LMC29293

| HJD (days-2450000) | $V_1$ km s$^{-1}$ | $V_2$ km s$^{-1}$ | Instrument |
|---|---|---|---|
| 6571.78391 | 276.215(200) | 322.550(214) | HARPS |
| 6578.82250 | 266.264(238) | 332.564(236) | HARPS |
| 6637.81235 | 331.213(142) | 271.716(148) | MIKE-Red |
| 6638.69816 | 333.351(136) | 271.308(124) | MIKE-Red |
| 6666.60395 | 328.940(128) | 272.911(132) | MIKE-Red |
| 7018.76308 | 341.798(146) | 264.069(142) | MIKE-Red |
| 7019.73177 | 340.750(138) | 262.992(134) | MIKE-Red |
| 7032.77143 | 334.658(136) | 270.047(136) | MIKE-Red |
| 7384.78486 | 338.589(176) | 265.287(170) | MIKE-Red |
| 8062.74807 | 261.668(104) | 336.736(104) | UVES |
| 8083.79085 | 272.044(106) | 326.160(110) | UVES |

**Notes.** RV are corrected by instrumental shift as described in section 3

## 4. Reddening and initial temperature estimation

For an initial temperature estimation of the systems, we used the infrared color $(V − K)$ corrected for the interstellar extinction. To get the intrinsic colors of the systems we estimated the interstellar extinction from the LMC reddening maps of Skowron

**Table 3.** Radial velocity measurements for the object LMC25304

| HJD (days-2450000) | $V_1$ (km s$^{-1}$) | $V_2$ (km s$^{-1}$) | Instrument |
|---|---|---|---|
| 5471.85308 | 253.823(196) | 306.291(194) | HARPS |
| 5558.68616 | 314.247(140) | 245.564(132) | MIKE-Red |
| 6309.75917 | 312.845(136) | 248.319(132) | MIKE-Red |
| 6577.82806 | 268.844(212) | 291.563(196) | HARPS |
| 6605.78760 | 251.803(200) | 308.494(210) | HARPS |
| 6637.62577 | 253.556(142) | 307.284(144) | MIKE-Red |
| 6638.54787 | 253.959(138) | 306.875(142) | MIKE-Red |
| 6932.77813 | 303.622(170) | 255.942(150) | MIKE-Red |
| 7004.79740 | 250.794(270) | 310.935(322) | HARPS |
| 7005.76097 | 250.254(228) | 311.053(238) | HARPS |
| 7018.80218 | 250.093(130) | 310.057(118) | MIKE-Red |
| 7030.71879 | 255.519(182) | 304.929(192) | HARPS |
| 7032.80173 | 256.927(136) | 304.281(128) | MIKE-Red |
| 7210.93242 | 250.177(140) | 311.073(150) | MIKE-Red |
| 7295.86523 | 318.201(150) | 241.807(138) | MIKE-Red |
| 8100.82997 | 302.754(100) | 257.272(100) | UVES |

**Notes.** RV are corrected by instrumental shift as described in section 3

et al. (2021) and Chen et al. (2022). Independently we derived reddenings from the sodium doublet NaI (5889.951, 5895.924 Å) using calibrations provided by Munari & Zwitter (1997) and Poznanski et al. (2012). We could only detect multicomponent absorption due to the Milky Way's interestellar material (ISM), while any absorption caused by the LMC's ISM was below the detection limit. Table 5 summarizes the determination of reddenings for our three systems where mean values are adopted as final interstellar extinction estimates.

In order to convert the reddening $E(B−V)$ into $E(V−K)$ we used the following relation from Savage & Mathis (1979):

$$E(V − K) = 2.72 E(B − V) \quad (1)$$

We used several $(V − K)$ color–effective temperature calibrations to estimate the main effective temperature of our targets, such as di Benedetto (1998), Tokunaga (2000), Houdashelt et al. (2000), Ramírez & Meléndez (2005), González Hernández & Bonifacio (2009), Casagrande et al. (2010) and Worthey & Lee (2011). We average estimates to obtain the initial "mean"

**Table 4.** Radial velocity measurements for the object LMC554.

| HJD (days-2450000) | $V_1$ km s$^{-1}$ | $V_2$ km s$^{-1}$ | Instrument |
|---|---|---|---|
| 6577.78370 | 313.085(250) | 267.055(164) | HARPS |
| 6579.80882 | 316.257(192) | 264.721(154) | HARPS |
| 6666.62651 | 252.511(190) | 329.268(166) | MIKE-Red |
| 6668.70476 | 252.542(196) | 327.753(168) | MIKE-Red |
| 6674.73038 | 256.464(180) | 324.120(233) | MIKE-Red |
| 6877.89612 | 321.478(304) | 259.192(288) | HARPS |
| 6931.81932 | 251.390(214) | 328.158(176) | MIKE-Red |
| 6932.87469 | 251.572(238) | 328.348(176) | MIKE-Red |
| 7018.82150 | 310.478(214) | 269.641(188) | MIKE-Red |
| 7047.61410 | 261.168(204) | 317.779(198) | MIKE-Red |
| 7270.90438 | 326.119(168) | 253.856(146) | UVES |
| 7383.79637 | 322.235(208) | 258.677(164) | MIKE-Red |
| 7384.80366 | 322.960(186) | 257.588(194) | MIKE-Red |
| 7403.64048 | 326.858(226) | 254.246(180) | MIKE-Red |
| 7404.66077 | 326.560(192) | 254.574(168) | MIKE-Red |
| 7658.82088 | 328.018(172) | 252.843(160) | UVES |
| 7728.74650 | 251.771(188) | 328.855(150) | UVES |
| 8062.69871 | 328.682(178) | 252.072(156) | UVES |

**Notes.** RV are corrected by instrumental shift as described in section 3

effective temperature for each system. In the case of LMC29293 we obtain $T = 5266K$, for the system LMC25304 we obtain $T = 4858K$ and for the system LMC554 we obtain $T = 5325K$.

## 5. Modeling approach and preliminary solution

In order to determine the physical characteristics of the systems we used the fitting routine implemented in the Wilson-Devinney (hereafter WD) program (Wilson & Devinney 1971; Wilson 1979, 1990; Van Hamme & Wilson 2007), version 2007. This routine allows to simultaneously fit light and radial velocity curves. The WD program utilizes a differential correction algorithm to find a set of parameters that describe observables. We adjusted simultaneously the I-band and the RV curves for each system. Mode 2 of the program was used as it was designed specifically for DEBs, and we set `IPB=0`, making luminosity and temperature coupled for the analysis. The albedo for both components was set to 0.5 and the gravity brightening $\beta$ to 0.32, i.e. standard values for stars with convective envelopes (Lucy 1967). In all of our modeling the logarithmic limb darkening law (Klinglesmith & Sobieski 1970) was used with automatically updated coefficients: `LD1=LD2=-2` as well as the stellar atmosphere formulation `IFAT1=IFAT2=1`. The applied proximity and eclipse effects to the radial velocity curves by setting `ICOR1=ICOR2=1`. We set the grid integer size as `N1=N2=60` to describe the size of the grid at the surface of the stars, and the noise parameter `NOISE=1` for observational scatter scaling with the square root of the light level. These settings are intended to optimize the interpretation of DEB data.

### 5.1. Initial parameters

The orbital periods and times of primary eclipse minimum were adopted from Pawlak et al. (2016) in the case of LMC29293 and LMC25304. We verified these values using the phase dispersion minimization (PDM) in IRAF[1] (Stellingwerf (1978)). For the system LMC554 the ephemeris was calculated using PDM. The parameters from RaveSpan RV fit (Section 3) were set as the initial orbital parameters for the fitting. The temperature of the primary star was set as the value from the color-effective temperature relations as described in Section 4. The metallicity was assumed to be the average metallicity for the LMC (Section 3).

### 5.2. Modelling strategy

The following parameters were adjusted simultaneously: the eccentricity $e$, the semi-major axis $a$, the orbital inclination $i$, the argument of periapsis $\omega$, the surface potentials $\Omega_1$ and $\Omega_2$, the systemic velocity $\gamma$ separately for each component, the mass ratio $q = M_2/M_1$, the effective temperature of the secondary component $T_2$, the orbital period $P$, the phase shift $\phi$, the light of primary component in given filter $L_1$ and also the third light $l_3$. The best solution was identified by a comparison of the resulting $\chi$-squared value calculated by comparing the model to the observed light curves.

During the light curve modeling it turned out that the radii of components, $r_1$ and $r_1$ are quite strongly correlated for all three systems. We employed the spectroscopic light ratio in $V$-band to break the degeneracy between the parameters. The model $V$-band ratios were extrapolated using the LC module of the WD code. After several unconstrained iterations a solution was found only for LMC25304 in which the model and observed spectroscopic light ratios agree within uncertainties. However, for LMC29293 and LMC554 we had to fix the potential of the secondary component in order to obtain an agreement with the spectroscopic light ratio.

After reaching a satisfactory solution new light ratios were used to recalculate the effective temperature of the components from the color-effective temperature relations listed in Section 4. After that, the WD models were recalculated once again with the new temperatures.

The presence of a third light $l_3$, which might suggest the existence of an additional body contributing to the total luminosity of the systems, was checked for all systems. We derived very low or negative values of $l_3$. Subsequently, we set $l_3 = 0$ in the case of LMC25304 and LMC554. In the case of LMC29293 we estimated a very weak contribution of third light.

## 6. Atmospheric parameters

### 6.1. Spectral decomposition

To identify atmospheric features of our stars from their spectra, such as metallicity, projected rotational velocities, microturbulence and effective temperatures, the composed spectra, with contributions from both stars, must be disentangled. To this end, used the González & Levato (2006) prescription implemented in RaveSpan. Only spectra listed as MIKE-Red in tables 2, 3 and 4 were used, being the ones with better SNR. The disentangled spectra are the result of an iterative process consisting of using the computed spectra of one of the stars, to calculate the other, and vice versa, subtracting the lines of one of the components with respect to its Doppler shift from the composed spectra at different phases.

---

[1] IRAF is distributed by the National Optical Astronomy Observatories, which are operated by the Association of Universities for Research in Astronomy, Inc., under a cooperative agreement with the National Science Foundation.

**Table 5.** Reddening estimates, E(B-V)

| Object | Map S21[a] | Map C22[b] | NaI D1 | Adopted $E(B-V)$ |
|---|---|---|---|---|
| LMC25304 | 0.082±0.028 | 0.079 | 0.068±0.015 | 0.076±0.015 |
| LMC29293 | 0.065±0.049 | 0.063 | 0.038±0.015 | 0.055±0.020 |
| LMC554   | 0.087±0.036 | 0.135 | 0.060±0.012 | 0.085±0.031 |

**Notes.** [a] (Skowron et al. 2021) [b] (Chen et al. 2022) - no uncertainties given.

**Table 6.** Results of GSSP atmospheric analysis.

| | LMC29293 | | LMC25304 | | LMC554 | |
|---|---|---|---|---|---|---|
| | Primary | Secondary | Primary | Secondary | Primary | Secondary |
| $T_{\text{eff}}$ (K) | 5271±190 | 5125±272 | 5155±222 | 5080±186 | 5517±396 | 5415±301 |
| [M/H] (dex) | -0.23±0.17 | -0.49±0.26 | -0.58±0.24 | -0.62±0.2 | -0.35±0.22 | -0.55±0.25 |
| $\xi$ (km s$^{-1}$) | 3.55±1.25 | 2.37±1.28 | 2.16±0.97 | 1.81±0.73 | 2.6±0.88 | 2.01±1.2 |
| $\zeta$ (km s$^{-1}$)[a] | 4.28 | 4.32 | 4.52 | 4.39 | 4.98 | 4.97 |
| $v \sin i$ (km s$^{-1}$) | $11.548^{+3.591}_{-4.23}$ | $9.371^{+3.88}_{-4.35}$ | 7.538±3.3 | 8.368±2.73 | 7.77±3.9 | 8.43±3.8 |

**Notes.** Log(g) was adopted from WD models and established as a settled parameter for the estimation of atmospheric quantities. a) Values presented for the macroturbulence velocity $\zeta$ were calculated using the equations from Section 6.2

### 6.2. Macroturbulence velocity determination

In order to derive the atmospheric parameters from the decomposed spectra of our systems we made some assumptions. The spectral SNR ~ 10 for our systems, due to the faintness of sources, thus this level of noise might induce doubtful determination of some parameters. To mitigate this effect we fixed the surface gravity log $g$ to a dynamical value derived from initial solutions found with the WD code. Regarding the macroturbulence velocity $\zeta$ it is difficult or even impossible to infer its value from low SNR spectra because of a strong correlation with the projected rotation velocity $v \sin i$. Thus we estimate the $\zeta$ from the following set of equations given by (Hekker & Meléndez (2007), Massarotti et al. (2008) and Takeda et al. (2008))

$$\zeta = 4.3 - 0.67 \log g \quad (2)$$
$$\zeta = 0.00195 T_{\text{eff}} - 3.953 \quad (3)$$
$$\log \zeta = 3.50 \log T_{\text{eff}} + 0.25 \log L/L_\odot - 12.97 \quad (4)$$

We take an average of the three above estimates as the value of $\zeta$. Those values were fixed during subsequent analysis of the decomposed spectra. The calculated values for macroturbulence of each star are listed in Table 6

### 6.3. Analysis with the GSSP code

A set of the atmospheric parameters such as the microturbulence velocity $\xi$ in the chromosphere of the stars, the effective temperature $T_{\text{eff}}$, the projected rotational velocity $v \sin i$, and the metallicity [M/H] were determined using the Grid Search in Stellar Parameters software (GSSP) published by Tkachenko (2015). This software uses a grid of those parameters and fits the best synthetic spectra to the observed data. The advantage of GSSP is that it is possible to fit simultaneously all aforementioned parameters for both stellar components within a binary configuration. The software computes synthetic spectra using the SynthV LTE-based radiative transfer code (Tsymbal 1996). It also uses a grid of atmosphere models pre-computed with LLmodels code (Shulyak et al. 2004), and also with the Model Atmospheres in a Radiative and Convective Scheme (MARCS), which is a grid of models for late-type stars (Gustafsson et al. 2008).

The general idea is to look for a better fit to the data around the selected input parameter values and to adopt the best set of parameters by calculating the Chi-square of residuals between the synthetic and the observed spectrum. The spectral region we analysed was in the range of 5985 - 6787 Å, since in this spectral region the level of noise is smaller relative to the bluer spectral regions. We also avoid regions with telluric lines listed by Kurucz (2005), such as $O_2$ around 6300 Å and $H_2O$ lines around 6500 Å. The result of our analysis is presented in Table 6

### 7. Adopted solution and calculation of uncertainties

The adopted solution is the combination of the results obtained with the WD code and the spectral analysis performed with GSSP. The final effective temperatures were estimated using the previously mentioned color-effective temperature relations in Section 4, applying the light ratios obtained from our WD models to disentangle the individual contribution of each star to the total observed magnitude in a given band. We calculated the arithmetic mean temperature with the GSSP temperature and the ones from the method mentioned above.

The statistical errors of the parameters were calculated using the JKTEBOP code (e.g. Southworth et al. (2005), Southworth et al. (2004)). JKTEBOP is a code to model and analyze detached binary systems from their light and radial velocity curves. The code is based on the EBOP code written by Popper & Etzel (1981) (see also Etzel (1981)). JKTEBOP is implemented with a Monte Carlo (MC) algorithm for uncertainty estimation.

As an input for JKTEBOP MC analysis, we used the WD model parameters derived in Section 5. We assumed that the parameters given by the WD model are a proper representation for the JKTEBOP code, which follows from the fact that all systems are well detached. The MC analysis performed with JKTEBOP works as follows: parameters adjusted are independently perturbed by small values and compared with the phased observed light curve. The resulting Corner plots of the analysis are shown in Appendix A. The outcomes of WD and JKTEBOP were compared in order to check if the parameters were in agreement with each other. The solution adopted is a combination of the parameters from WD and uncertainties derived with JKTEBOP presented in Table 7.

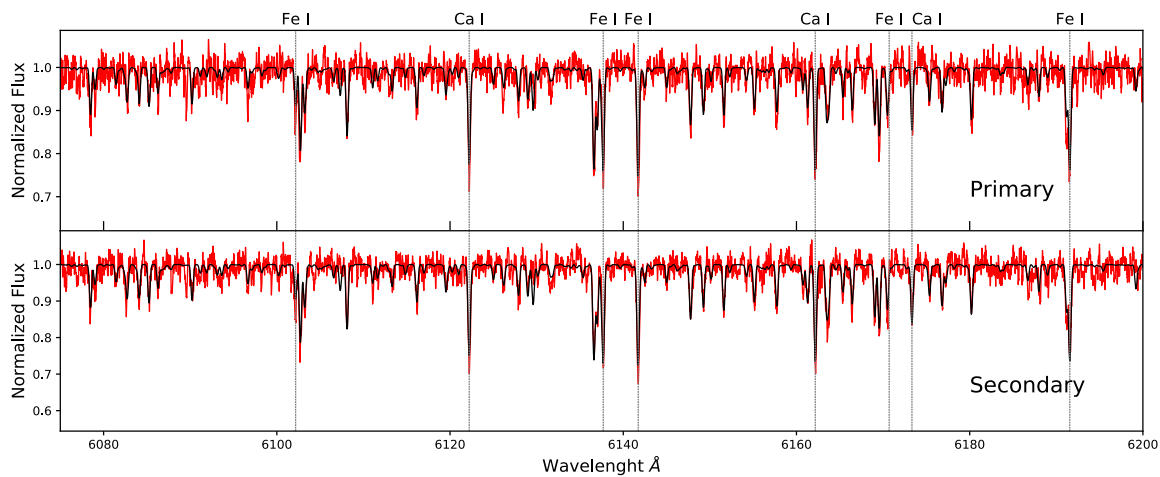

**Fig. 2.** 6080 - 6200 Å spectral region of the object LMC25304. Red lines are the disentangled spectra from MIKE-red observations. Black lines are the synthetic spectra generated using GSSP binary mode.

The physical parameters were derived using a set of astrophysical constants advocated by Prša et al. (2016) and they are summarized in Table 8. The spectral types were estimated based on the surface temperatures and gravities according to Table 1 from Alonso et al. (1999). Below we described individual systems in more detail.

### 7.1. OGLE-LMC-ECL-29293

LMC29293 is a detached binary composed of two very similar late-type giant stars in terms of temperature, radii, luminosity, and masses. Both stars were classified as G4III and the system is the most metal-rich in our sample ([M/H]= $-0.37 \pm 0.15$ dex). The secondary is the slightly hotter star of the system, while also being the more massive component. The above description is a bit discrepant with the GSSP results which prefer the primary star to be slightly hotter, however, it might be explained by the relatively low SNR of the analyzed spectra.

The eccentricity of the system is very low – see Fig. 3 – we obtained $e = 0.002 \pm 0.003$ and we fixed $e = 0$ during subsequent analysis. A small amount of third light was detected but its influence on our solution is insignificant and we also fixed $\ell_3 = 0$. The light curve analysis showed that an unconstrained WD solution has a strong correlation between the radii of the components and it was practically impossible to make the parameters converge. We used the spectroscopic light ratio to restrict the light ratio of components by fixing the secondary surface potential. In this way we were able to break the degeneracy of the photometric parameters.

This system shows the most unequal mass components of our sample; the remaining two systems have a mass ratio very close to unity, which concurs with the physical similarity of their components. In the case of LMC29293 we have two stars of no-tably similar external characteristics yet they differ significantly in masses. We address this problem in Section 8.1.

### 7.2. OGLE-LMC-ECL-25304

The object LMC25304 is an eclipsing binary composed of two similar late-type giant stars. We estimated their spectral types as G4III+G5III. The eclipses are deep and almost central, however, no totality phase can be seen – see Fig. 4. The components have the largest masses and radii in our sample, alongside the widest separation and significant eccentricity. Additionally, this system has a mass ratio below 1, meaning the hotter primary component is slightly more massive than its companion.

During the light curve analysis, we found a small negative value of third light in the I band ($\ell_3 = -0.0016(89)$), therefore we set $\ell_3 = 0$ in our adopted solution. The orbital inclination $i$ of the orbit is so close to 90 degrees that it causes a numerical instability while solving with the WD code. Using Monte Carlo simulations with the JKTEBOP code we restricted a range of possible inclinations (see - Appendix A) and during subsequent analysis we fixed the orbital inclination at $i = 89.775$ deg.

This is the only system in our sample where an unconstrained light curve solution leads to a light ratio that is fully consistent with the spectroscopic light ratio. Subsequently, we did not fix any other parameter during the analysis. The adopted solution does not show any significant systematic residuals. Interestingly although the system is the brightest in our sample its I-band light curve solution shows the largest *rms*. We searched for periodicity in residuals using the Phase Dispersion Minimization (PMD) method Stellingwerf (1978) and the Lomb-Scarle periodogram Lomb (1976); Scargle (1982). Both methods returned strongest power peaks at frequencies corresponding to one year and one day and their harmonics (Fig. 5). The former is caused by systematic seasonal brightness trends due to airmass changes, while the latter arises from a basic 1-day OGLE cadence. However, both effects combined do not explain the residual spread. No other significant frequency peaks were detected; thus, probably one or both components of LMC25304 exhibit some short-term erratic variability which increases residuals.

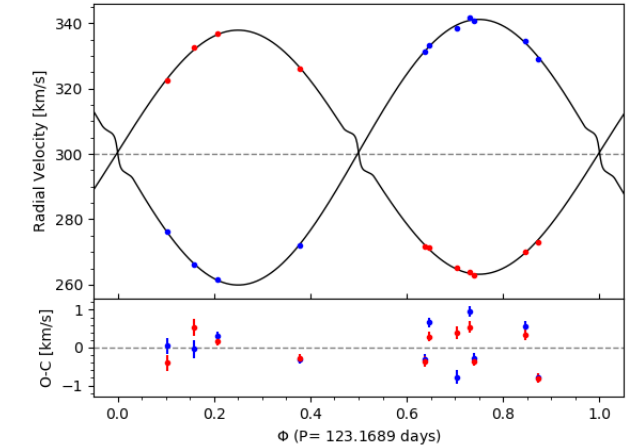

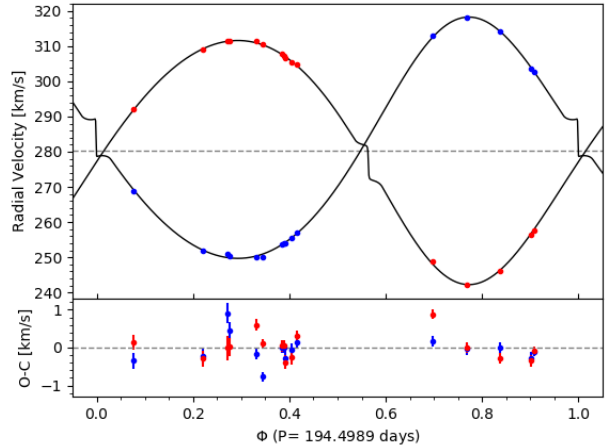

**Fig. 3.** The radial velocity data and its binary solution (upper panel), and the photometric light curve with its solution (lower panel) obtained with the WD code for the system LMC29293. The smaller lower panel shows the residuals of our adopted solutions..

**Fig. 4.** Same as Fig. 3, but for the binary LMC25304.

### 7.3. LMC554.19.81

This system is composed of the hottest stars in the sample. We estimated their spectral types as G2III+G2III. Like in both previous systems, the eclipses are partial, deep, and well separated. The OGLE $I$-band light curve is of good quality and the smallest $rms$ in the sample – see Fig. 6. No significant out-of-eclipse variability could be noticed. No third light was detected, however, we found a small, yet significant eccentricity which reduced residuals of the WD solution to the light and radial velocity curves.

As in the case of LMC29293 we found a strong correlation between component radii during analysis with the WD code. The correlation was not so strong like in LMC-29293 but we had to constrain a model light ratio by using the spectroscopic value and fixing the secondary surface potential. In general, the errors on parameters we obtained for this system were the smallest in the sample.

## 8. Evolutionary stage

From our solutions, we obtained very well determined masses, radii, luminosities, and effective temperatures for each analyzed star. We also obtained an estimation of the individual abundances of the stars from our spectroscopic analysis. Implementing those values, we were able to construct isochrones for our sources with the aim of determining their ages. We used the MESA Isochrones and Stellar Tracks (MIST) library (see Choi et al. (2016) and Dotter (2016) for details) computed using the Modules for Experiments in Stellar Astrophysics (MESA) code (Paxton et al. (2011)). The MIST models are scaled to the abundance of the sun and evolve the stars from the ZAMS to later stages of stellar evolution.

For the analysis, we selected a grid of isochrones that cover the range of log(Age) from 8 to 9 in steps of 0.005 dex and corresponding to the spectroscopic metallicities derived with GSSP. We compare the position of our targets in the $\log(T_{\rm eff})$ vs $\log L/L_\odot$ plane with the isochrones and check for consistency with masses and radii of each component. Since our stars seem to be passing through the same evolutionary stage, it is expected that the formation of both components happened at a similar time with identical initial composition.

For a given age, we select three isochrones that differ in abundance corresponding to the mean, lower, and upper values for the metallicity of a system given in Table 8. By varying the isochrone age we searched for the age that minimizes the distance between the theoretically predicted mass and the observed one, while simultaneously satisfying the position of our stars

**Table 7.** Orbital and photometric parameters from the adopted solution using WD and JKTEBOP codes.

| | LMC29293 | LMC25304 | LMC554 |
|---|---|---|---|
| Orbital Period P (days)$^a$ | 123.1689± 0.0038 | 194.4989 ± 0.0041 | 133.1725 ± 0.0009 |
| $T_0$ (JD-2450000)$^a$ | 6436.018 ± 0.012 | 7146.410 ± 0.030 | 6497.7498 ± 0.0039 |
| Inclination $i$ (deg)$^a$ | 87.792 ±0.051 | 89.775 (fixed) | 88.816 ± 0.036 |
| Semi-major axis ($R_\odot$)$^b$ | 189.90 ± 0.68 | 263.68± 0.79 | 203.28 ± 0.41 |
| Eccentricity $e^b$ | 0 (fixed) | 0.1074 ±0.0010 | 0.004 ± 0.002 |
| Longitud of periastron $\omega$ (deg)$^b$ | - | 342.3 ±1.6 | 88.3 ± 3.7 |
| Systemic Velocity $\gamma_1$ (km s$^{-1}$)$^b$ | 300.51±0.17 | 280.46 ± 0.09 | 290.24 ±0.06 |
| $\gamma_2 - \gamma_1$ (km s$^{-1}$) | 0.08 | −0.44 | −0.02 |
| RV Semi-amplitude $K_1$ (km s$^{-1}$)$^b$ | 40.62 ± 0.22 | 34.29 ± 0.14 | 38.72 ± 0.11 |
| RV Semi-amplitude $K_2$ (km s$^{-1}$)$^b$ | 37.33 ± 0.17 | 34.69 ±0.13 | 38.49 ± 0.11 |
| Mass Ratio **q**$^b$ | 1.088 ±0.008 | 0.988 ±0.005 | 1.006 ±0.004 |
| Surface Potential $\Omega_1^b$ | 10.860 ± 0.144 | 12.264 ± 0.121 | 11.783 ± 0.079 |
| Surface Potential $\Omega_2^b$ | 11.660 (fixed) | 11.509 ± 0.135 | 11.410 (fixed) |
| Sum of Fractional Radii $\frac{R_P+R_S}{A}$$^a$ | 0.2041 ± 0.0018 | 0.1849 ± 0.0018 | 0.1896 ± 0.0016 |
| Ratio of Radii **k** = $\frac{R_S}{R_P}$$^a$ | 0.993±0.025 | 1.062± 0.030 | 1.041 ± 0.021 |
| Primary's Temperature $T_1$ (K) | 5170 (fixed) | 5150 (fixed) | 5450 (fixed) |
| Secondary's Temperature $T_2$(K)$^b$ | 5215 ± 12 | 5053±20 | 5421± 7 |
| Light Ratio $(\frac{L_S}{L_P})_I^b$ | 1.019± 0.048 | 1.048 ± 0.035 | 1.064 ± 0.032 |
| $(\frac{L_S}{L_P})_V^c$ | 1.033 | 1.018 | 1.055 |
| Third light $\ell_{3\,I}^b$ | 0 (fixed) | 0 (fixed) | 0 (fixed) |
| rms$_{RV1}$ (m s$^{-1}$) | 548 | 349 | 353 |
| rms$_{RV2}$ (m s$^{-1}$) | 439 | 318 | 365 |
| rms Ic (mmag) | 9.0 | 13.7 | 6.6 |

**Notes.** Parameters marked as fixed were decided to be settled in order to avoid a degeneracy in model calculations.
a) Calculated with WD code with uncertainties from the Monte Carlo algorithm with JKTEBOP.
b) Uncertainties calculated with differential correction from WD code.
c) Extrapolated value from WD solution.

in the Mass vs. Radius, Mass vs Luminosity, and Mass vs $T_{\text{eff}}$ planes. The consistency of our parameters with the theoretical predictions for a certain age is shown in the figures in Appendix B.

For all three cases, we noticed that our systems must be placed at the so-called blue loop when stars begin to ignite helium in their helium-rich cores through the triple $\alpha$ process. This makes a star move to a bluer region of the HRD, getting hotter for a period, and then cooling down again before entering the asymptotic giant branch (AGB). Our estimated ages are presented in Table 8.

### 8.1. OGLE-LMC-ECL-29293

System LMC29293 presents the most complex scenario for evolutionary analysis within our sample. As previously mentioned, the masses of these stars are the most different from each other among the three systems. On the other hand, both stars should be in the same evolutionary stage since they are very close to each other in the HR diagram – see Fig. 7.

The difference in masses indicates that the secondary component, i.e. the more massive one, should be slightly more evolved, already approaching the AGB, while the primary should still be getting hotter and should just enter its blue loop after passing through the helium flash. According to the proposed evolutionary stage, our system is composed of stars with ages in the range of 8.502 < log(Age) < 8.558, giving a mean age of around 337 Myr. When evolutionary tracks are fitted separately for each component the best fit ages are different by 62±12 Myr assuming the same chemical composition.

Several reasons can be invoked to explain why it is not possible to simultaneously fit an isochrone for both stars. The system may be the result of a stellar capture may be a former hierarchical triple system in which one of the components is the result of a stellar merger. However, the observed similarity of components is difficult to explain in both scenarios without assuming some very intricate fine tuning of stellar parameters.

Table 8. Fundamental and physical properties derived from the adopted solution.

| | LMC29293 | | LMC25304 | | LMC554 | |
|---|---|---|---|---|---|---|
| | Primary | Secondary | Primary | Secondary | Primary | Secondary |
| Spectral Type | G4III | G4III | G4III | G5III | G2III | G2III |
| Mass ($M_\odot$) | 2.898±0.031 | 3.153±0.038 | 3.267±0.028 | 3.229±0.029 | 3.165±0.020 | 3.184±0.020 |
| Radii ($R_\odot$) | 19.43±0.31 | 19.30±0.31 | 23.62±0.42 | 25.10±0.43 | 18.86±0.26 | 19.64±0.26 |
| Surface gravity log $g$ (dex) | 2.323±0.015 | 2.365 ± 0.015 | 2.205±0.016 | 2.148±0.015 | 2.387 ±0.012 | 2.355±0.012 |
| Effective temp. $T_{eff}(K)$ | 5141±85 | 5186±86 | 5157±75 | 5060±74 | 5450±125 | 5420±125 |
| Luminosity ($L_\odot$) | 237±17 | 242±18 | 355±23 | 371±24 | 282±27 | 299±29 |
| Metallicity [M/H] (dex) | -0.37±0.15 | | -0.6±0.15 | | -0.45± 0.16 | |
| $M_{bol}$ (mag) | -1.196±0.077 | -1.219±0.080 | -1.635±0.070 | -1.683±0.070 | -1.385±0.103 | -1.449±0.105 |
| $V$ (mag) | 17.815±0.019 | 17.518±0.019 | 17.289±0.021 | 17.272±0.021 | 17.572±0.019 | 17.476±0.019 |
| $I_c$ (mag) | 16.731± 0.032 | 16.716 ± 0.032 | 16.302± 0.020 | 16.256 ± 0.020 | 16.808± 0.047 | 16.774 ± 0.046 |
| Age (Myr) | 337±21 | | 270±20 | | 285±6 | |
| Distance$_{phot}$ (kpc) | 49.094±1.572 | | 48.58±4.48 | | 51.65±2.04 | |

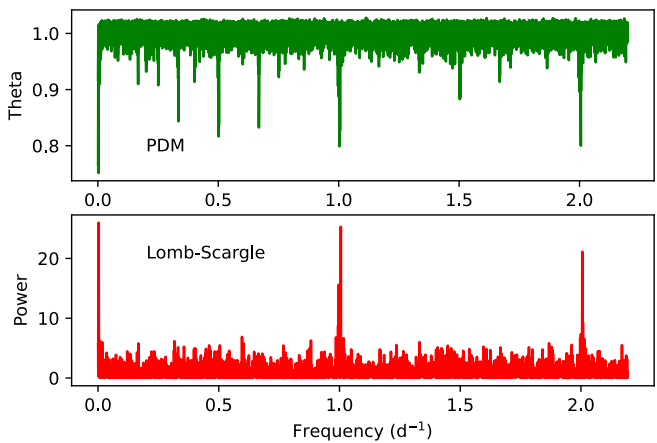

Fig. 5. The results of periodicity search in light residuals of LMC25304. The upper panel shows the PDM diagram and the lower panel shows the Lomb-Scargle periodogram.

Another possible explanation might be differences in the initial rotation rates of the stars, discrepant initial compositions, or interaction with another star that changes the context of the binary. Indeed the system's components show the most discrepant surface metallicities in the sample, however, uncertainties are also large and the difference is only slightly larger than $1\sigma$.

### 8.2. OGLE-LMC-ECL-25304

The comparison with evolutionary models suggests that both stars are in the same evolutionary stage – see Fig. 8. They have just entered their blue loops, getting slightly hotter as helium burning is happening at their center. The isochrone most consistent with observed values is the one corresponding to the adopted mean metallicity, [M/H] = −0.6 dex, and with an age of log(Age) = 8.416, i.e. 260 Myr. Both stars are about $2\sigma$ off their predicted position: the difference of their masses suggests a larger difference of surface temperatures than is actually observed.

### 8.3. LMC554.19.81

The best solution places the components of the system in the blue edge of the loop on the HR diagram – see Fig. 9. For metallicities close to the system's mean, i.e. [M/H]=−0.45 dex we obtain fully satisfactory fits and the stars nearly reach their maximum temperature. Both $1\sigma$ higher and lower metallicities (i.e. [M/H]=−0.6 and −0.3 dex) give significantly worse fits to observed position of components. We estimate the age of the system to be between 8.46 < log(Age) < 8.56, or approximately 280 Myr.

## 9. Photometric distances

Using the parameters established throughout this work, along with individual bolometric corrections from Alonso et al. (1999), we determined photometric distances by applying the following equation from Graczyk et al. (2017):

$$d_i(\text{pc}) = 3.360 \times 10^{-8} R_i T_i^2 10^{0.2(BC_i+V_i)} \quad (5)$$

with $i = 1, 2$ corresponds to the primary and secondary stars respectively. T is the effective temperature of each star in K, R is their stellar radius in units of solar radii and V is the intrinsic magnitude of the star corrected by extinction. After calculating the photometric distance to each star, we determined the distance to the binary system by averaging the distances of the two stars. The final photometric distances are shown in Table 8.

## 10. Conclusions

In this work, we derived fundamental physical quantities such as masses and radii for all our systems with precision better than 2%. All six stars belong to a population of yellow/red giant stars, and they are similar in terms of temperatures, radii, and masses.

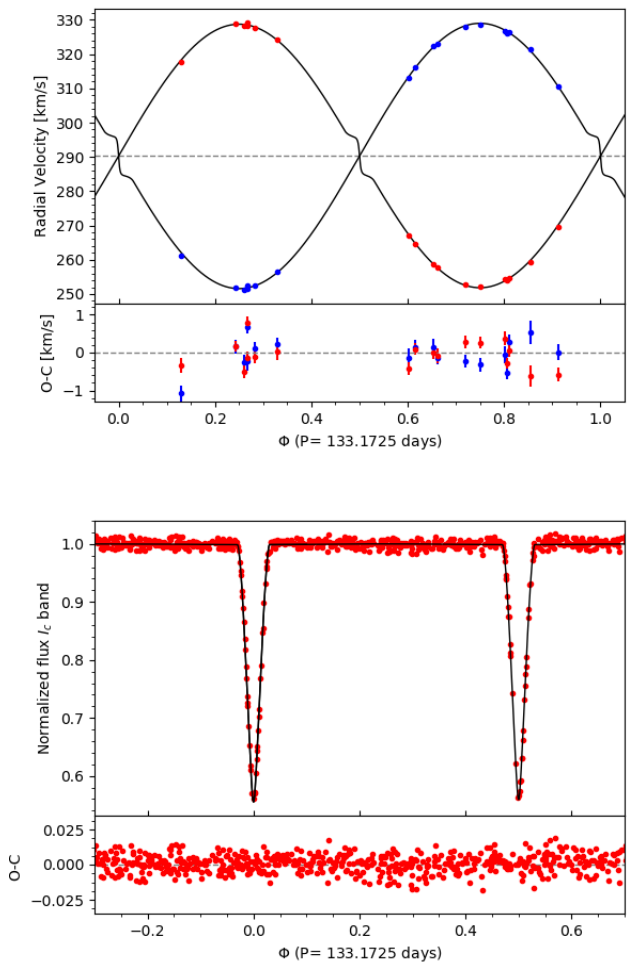

**Fig. 6.** Same as Fig. 3, but for the binary LMC554.

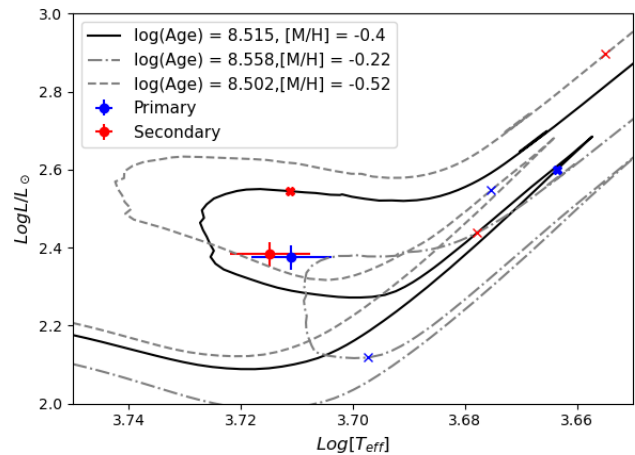

**Fig. 7.** The determination of age for the system LMC29293 with theoretical isochrones (black lines). Dashed gray lines correspond to $\pm 1\sigma$ uncertainty limits for metallicity from Table 8. Blue and red crosses mark the masses and luminosities of the primary and secondary components, respectively.

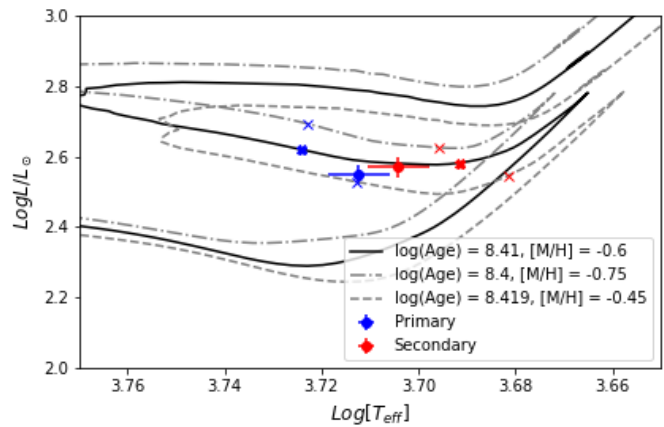

**Fig. 8.** Same as Figure 7, but for the system LMC25304.

As discussed in the previous section, the determination of the metal abundance with high precision of each star, a parameter with strong impact on the age determination, is essential for accurate determination of the evolutionary stage of the sources. For example, in the case of the system LMC29293, this is the main limiting factor, preventing us from a satisfactory explanation for the disagreement between the observed and predicted positions of the objects in the HR diagram.

The relatively low SNR of the decomposed spectra limits the precision of spectroscopically derived surface temperatures and metallicities. In this context, it was crucial to resort to other methods such as the temperature-color relations and the temperature ratio provided by the WD solution. Subsequently, we achieved temperature uncertainties better than 100 K for four stars and about 125 K for the remaining two.

This part of the analysis leads us to believe there are still opportunities to improve these measurements with additional spectroscopic observations. For example, the PLATO satellite will provide another valuable opportunity to enhance our comprehension of these sources. Once in operation, the PLATO field of view will contain about half of the Large Magellanic Cloud body (Nascimbeni et al. (2022)), offering a new opportunity to refine our models and confirm the possibility of intrinsic variability of the targets.

Our study adds three new objects to the DEBs with giant star components already studied in the LMC by the Araucaria Project. These new systems have on average larger distances from the LMC center than systems previously analysed by our team. In the next step we plan to determine distances to these sources using the Surface brightness - Color Relation (SBCR) method, which, in addition to 20 other similar DEBs in the LMC, will improve our knowledge of the distance to this galaxy as well as its structure and the metallicity distribution across it.

*Acknowledgements.* The research leading to these results has received funding from the European Research Council (ERC) under the Horizon 2020 program (grant No. 951549 - UniverScale), the Polish Ministry of Science and Higher Education (grant 2024/WK/02), and the Polish-French Marie Skłodowska-Curie and Pierre Curie Science Prize from the Foundation for Polish Science. G.R.G. acknowledges the financial support of the National Science Centre, Poland (NCN) under the project No. 2017/26/A/ST9/00446. This work has been co-funded by the National Science Centre, Poland, grant No. 2022/45/B/ST9/00243.

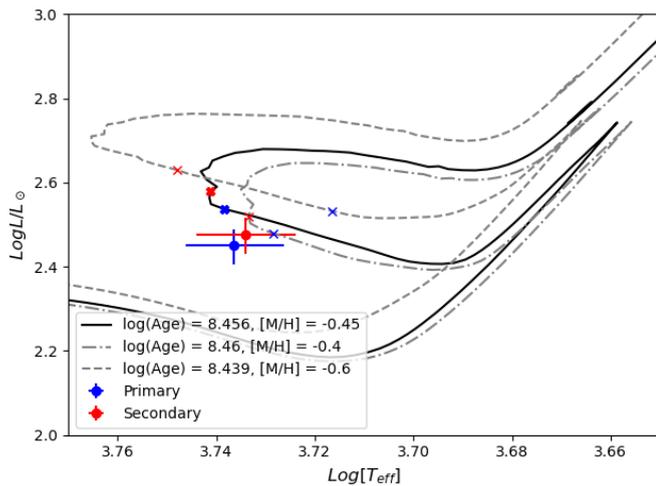

**Fig. 9.** Same as Figure 7, but for the system LMC554.

# References

Alonso, A., Arribas, S., & Martínez-Roger, C. 1999, A&AS, 140, 261
Bernstein, R., Shectman, S. A., Gunnels, S. M., et al. 2003, Proc. SPIE, 4841, 1694
Bonanos, A. Z., Stanek, K. Z., Kudritzki, R. P., et al. 2006, ApJ, 652, 313
Casagrande, L., Ramírez, I., Meléndez, J., et al. 2010, A&A, 512, A54
Chen, B.-Q., Guo, H.-L., Gao, J., et al. 2022, MNRAS, 511, 1317
Choi, J., Dotter, A., Conroy, C., et al. 2016, ApJ, 823, 102
Choudhury, S., de Grijs, R., Bekki, K., et al. 2021, MNRAS, 507, 4752
Claret, A. & Torres, G. 2019, ApJ, 876, 134
Coelho, P., Barbuy, B., Meléndez, J., et al. 2005, A&A, 443, 735
Cutri, R. M., Skrutskie, M. F., van Dyk, S., et al. 2012, VizieR Online Data Catalog, II/281
Dekker, H., D'Odorico, S., Kaufer, A., et al. 2000, Proc. SPIE, 4008, 534
del Burgo, C. & Allende Prieto, C. 2018, MNRAS, 479, 1953
di Benedetto, G. P. 1998, A&A, 339, 858
Dotter, A. 2016, ApJS, 222, 8
Eker, Z., Bakış, V., Bilir, S., et al. 2018, MNRAS, 479, 5491
Etzel, P. B. 1981, Photometric and Spectroscopic Binary Systems, 69, 111
Gaulme, P., McKeever, J., Jackiewicz, J., et al. 2016, ApJ, 832, 121
González Hernández, J. I. & Bonifacio, P. 2009, A&A, 497, 497
González, J. F. & Levato, H. 2006, A&A, 448, 283
Graczyk, D., Soszyński, I., Poleski, R., et al. 2011, Acta Astron., 61, 103
Graczyk, D., Pietrzyński, G., Thompson, I. B., et al. 2014, ApJ, 780, 59
Graczyk, D., Konorski, P., Pietrzyński, G., et al. 2017, ApJ, 837, 7
Graczyk, D., Pietrzyński, G., Thompson, I. B., et al. 2018, ApJ, 860, 1
Graczyk, D., Pietrzyński, G., Thompson, I. B., et al. 2020, ApJ, 904, 13
Graczyk, D., Pietrzyński, G., Galan, C., et al. 2021, A&A, 649, A109
Gustafsson, B., Edvardsson, B., Eriksson, K., et al. 2008, A&A, 486, 951
Hekker, S. & Meléndez, J. 2007, A&A, 475, 1003
Hełminiak, K. G., Konacki, M., Maehara, H., et al. 2019, MNRAS, 484, 451
Higl, J. & Weiss, A. 2017, A&A, 608, A62
Houdashelt, M. L., Bell, R. A., & Sweigart, A. V. 2000, AJ, 119, 1448
Klinglesmith, D. A. & Sobieski, S. 1970, AJ, 75, 175
Kato, D., Nagashima, C., Nagayama, T., et al. 2007, PASJ, 59, 615
Kruszewski, A. & Semeniuk, I. 1999, Acta Astron., 49, 561
Kurucz, R. L. 2005, Memorie della Societa Astronomica Italiana Supplementi, 8, 86
Lomb, N. R. 1976, Ap&SS, 39, 447
Lucy, L. B. 1967, ZAp, 65, 89
Masana, E., Jordi, C., & Ribas, I. 2006, A&A, 450, 735
Massarotti, A., Latham, D. W., Stefanik, R. P., et al. 2008, AJ, 135, 209
Mayor, M., Pepe, F., Queloz, D., et al. 2003, The Messenger, 114, 20
Munari, U. & Zwitter, T. 1997, A&A, 318, 269
Nascimbeni, V., Piotto, G., Börner, A., et al. 2022, A&A, 658, A31
Oh, S., Price-Whelan, A. M., Brewer, J. M., et al. 2018, ApJ, 854, 138
Paczynski, B. 1997, The Extragalactic Distance Scale, 273
Pawlak, M., Soszyński, I., Udalski, A., et al. 2016, Acta Astron., 66, 421
Paxton, B., Bildsten, L., Dotter, A., et al. 2011, ApJS, 192, 3
Pietrzyński, G., Thompson, I. B., Graczyk, D., et al. 2009, ApJ, 697, 862
Pietrzyński, G., Graczyk, D., Gieren, W., et al. 2013, Nature, 495, 76
Pietrzyński, G., Graczyk, D., Gallenne, A., et al. 2019, Nature, 567, 200
Pilecki, B., Graczyk, D., Pietrzyński, G., et al. 2013, MNRAS, 436, 953
Pojmanski, G. 1997, Acta Astron., 47, 467
Popper, D. M. & Etzel, P. B. 1981, AJ, 86, 102
Poznanski, D., Prochaska, J. X., & Bloom, J. S. 2012, MNRAS, 426, 1465
Prša, A., Harmanec, P., Torres, G., et al. 2016, AJ, 152, 41
Ramírez, I. & Meléndez, J. 2005, ApJ, 626, 446
Riess, A. G., Yuan, W., Macri, L. M., et al. 2022, ApJ, 934, L7
Rozyczka, M., Thompson, I. B., Dotter, A., et al. 2022, MNRAS, 517, 2485
Rucinski, S. 1999, Turkish Journal of Physics, 23, 271
Savage, B. D. & Mathis, J. S. 1979, ARA&A, 17, 73
Scargle, J. D. 1982, ApJ, 263, 835
Shulyak, D., Tsymbal, V., Ryabchikova, T., et al. 2004, A&A, 428, 993
Skowron, D. M., Skowron, J., Udalski, A., et al. 2021, ApJS, 252, 23
Southworth, J., Smalley, B., Maxted, P. F. L., et al. 2005, MNRAS, 363, 529
Southworth, J., Maxted, P. F. L., & Smalley, B. 2004, MNRAS, 351, 1277
Southworth, J. 2015, Living Together: Planets, Host Stars and Binaries, 496, 164
Stellingwerf, R. F. 1978, ApJ, 224, 953
Suchomska, K., Graczyk, D., Smolec, R., et al. 2015, MNRAS, 451, 651
Takeda, Y., Sato, B., & Murata, D. 2008, PASJ, 60, 781
Thompson, I. B., Kaluzny, J., Pych, W., et al. 2001, AJ, 121, 3089
Tkachenko, A. 2015, A&A, 581, A129
Tokunaga, A. T. 2000, Allen's Astrophysical Quantities, 143
Torres, G., Andersen, J., & Giménez, A. 2010, A&A Rev., 18, 67
Tsymbal, V. 1996, M.A.S.S., Model Atmospheres and Spectrum Synthesis, 108, 198
Udalski, A., Kubiak, M., & Szymanski, M. 1997, Acta Astron., 47, 319
Udalski, A. 2003, Acta Astron., 53, 291
Udalski, A., Soszynski, I., Szymanski, M. K., et al. 2008, Acta Astron., 58, 89
Udalski, A., Szymański, M. K., & Szymański, G. 2015, Acta Astron., 65, 1
Van Hamme, W. & Wilson, R. E. 2007, ApJ, 661, 1129
Vilardell, F., Ribas, I., Jordi, C., et al. 2010, A&A, 509, A70
Wilson, R. E. & Devinney, E. J. 1971, ApJ, 166, 605
Wilson, R. E. 1979, ApJ, 234, 1054
Wilson, R. E. 1990, ApJ, 356, 613
Worthey, G. & Lee, H.-. chul . 2011, ApJS, 193, 1
Zaritsky, D., Harris, J., Thompson, I. B., et al. 2004, AJ, 128, 1606

# Appendix A: Results of Monte Carlo algorithm over orbital and photometric parameters.

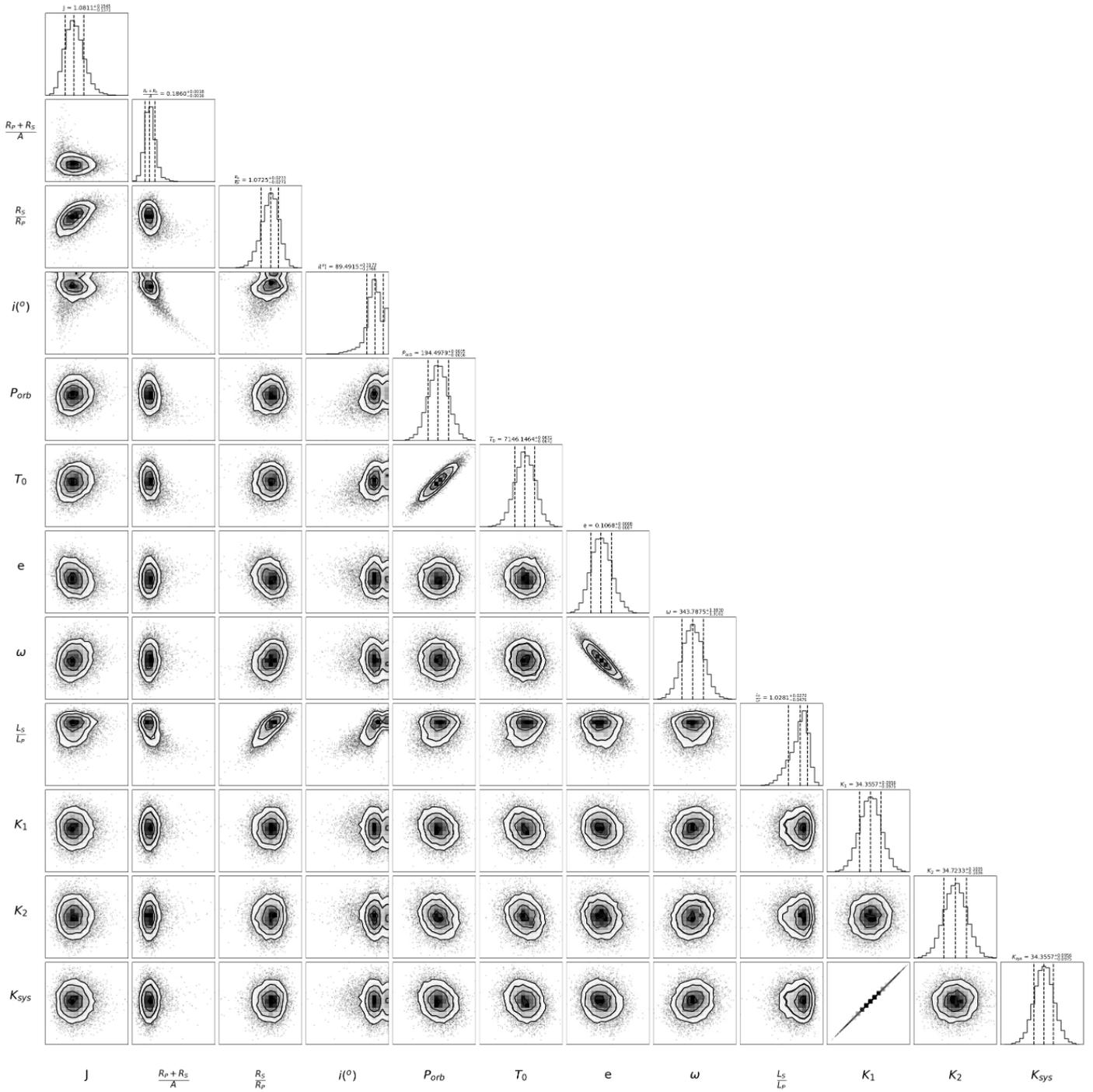

**Fig. A.1.** Corner plot of the perturbed parameters analyzed with Monte Carlo algorithm for the system LMC25304

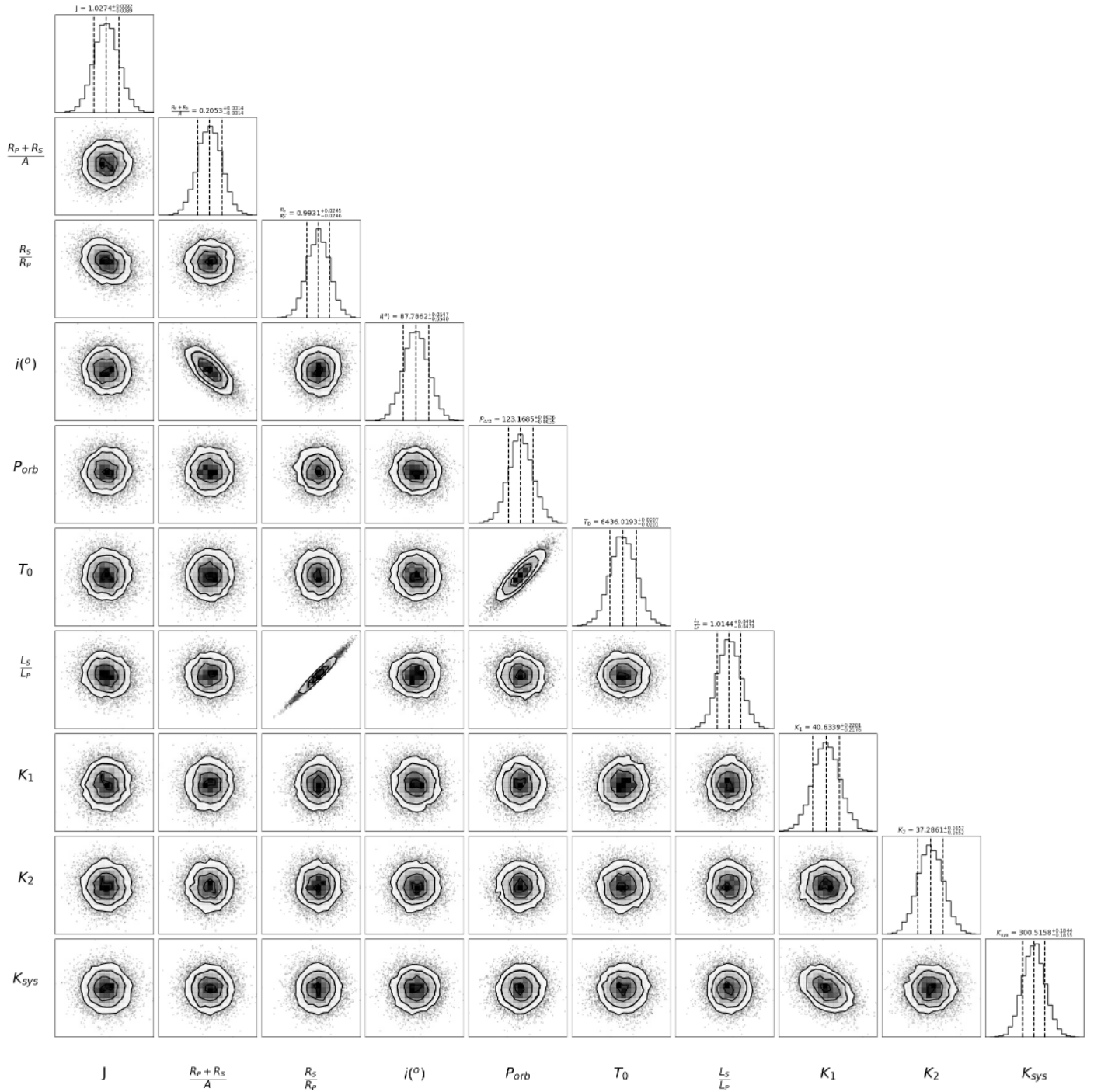

**Fig. A.2.** Corner plot of the perturbed parameters analyzed with Monte Carlo algorithm for the system LMC29293. The model satisfies the observations assuming a circular orbit, which means e=0. Therefore, we fixed e and was not included in the MC calculations.

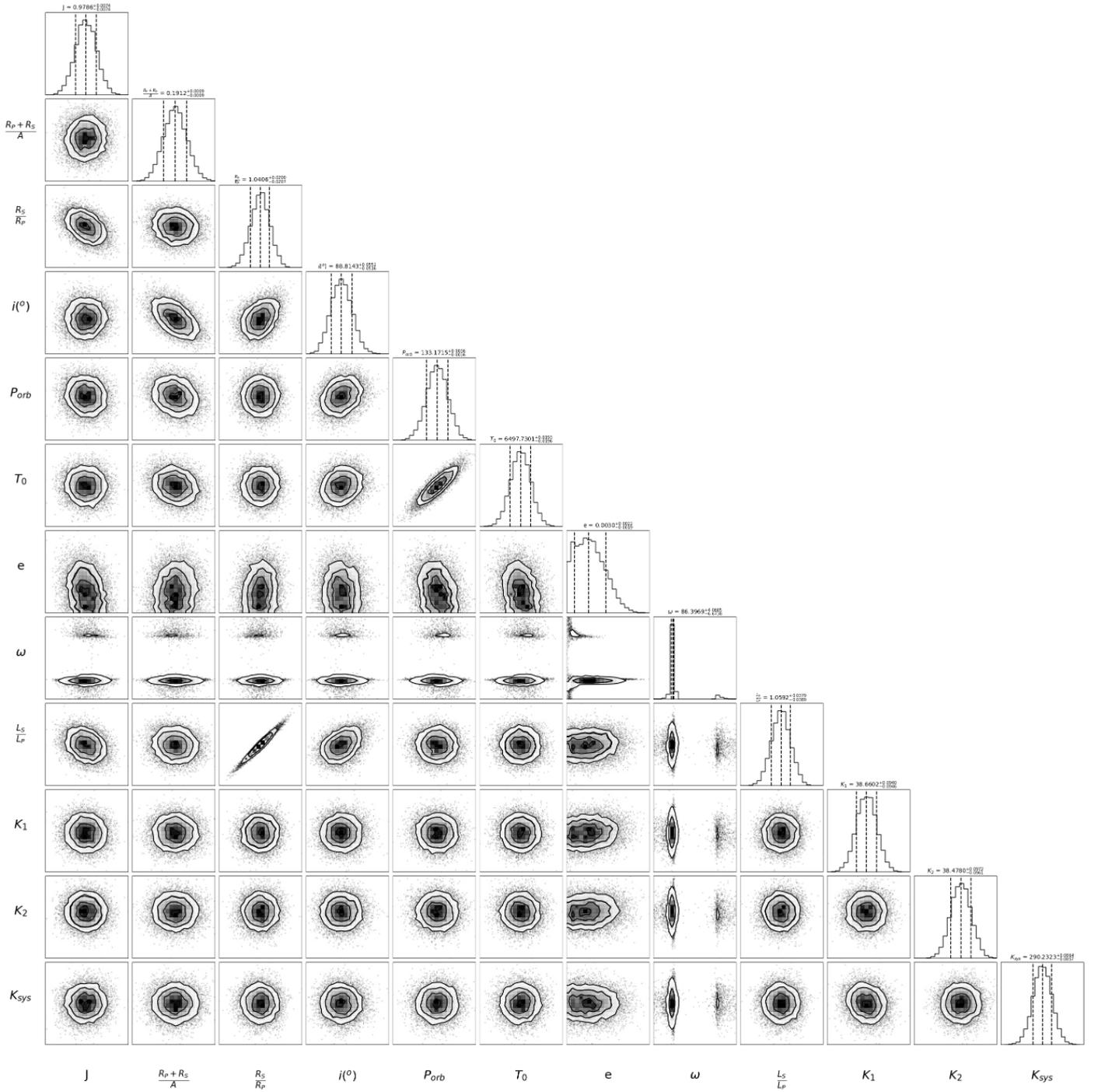

**Fig. A.3.** Corner plot of the perturbed parameters analyzed with Monte Carlo algorithm for the system LMC554.

# Appendix B: Comparison with MIST isochrones.

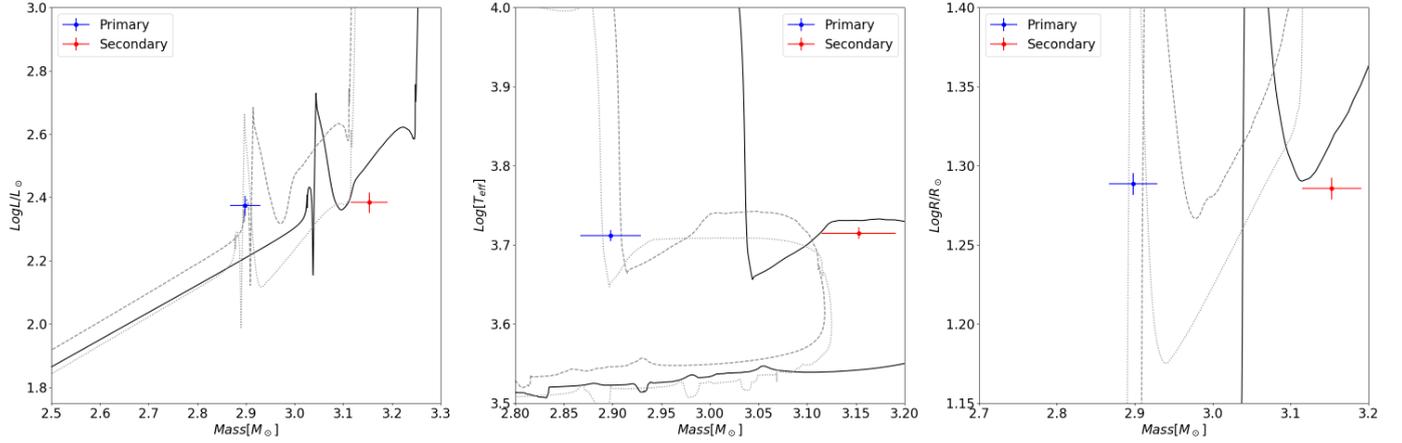

**Fig. B.1.** Mass vs luminosity, effective temperature, and radius diagrams. Solid line is the main metallicity of the system, and dashed lines represent the lower and upper limits of the determined metallicity for the system LMC29293.

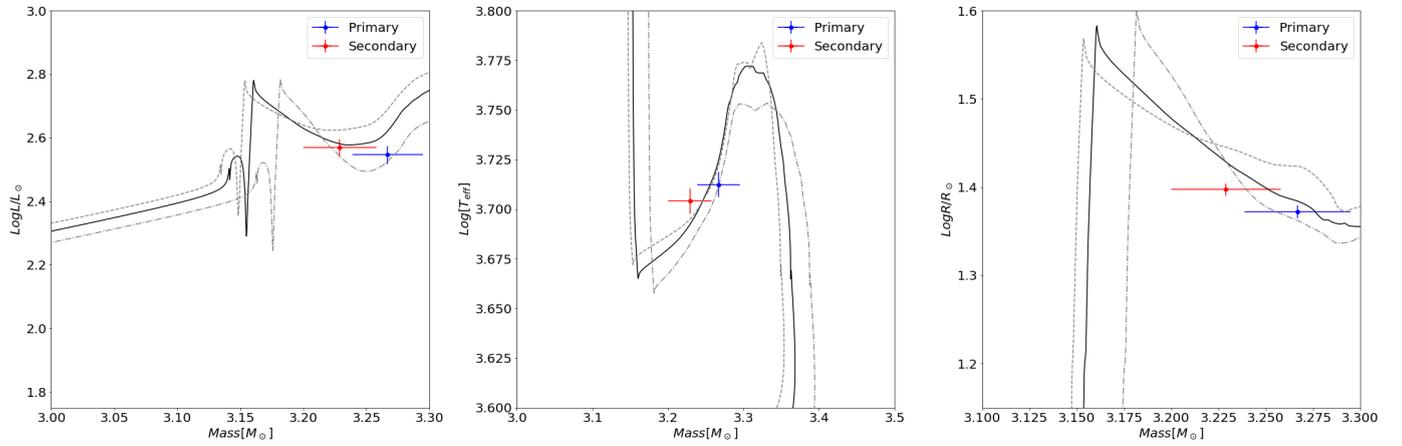

**Fig. B.2.** Same as Figure B.1 but for the system LMC25304.

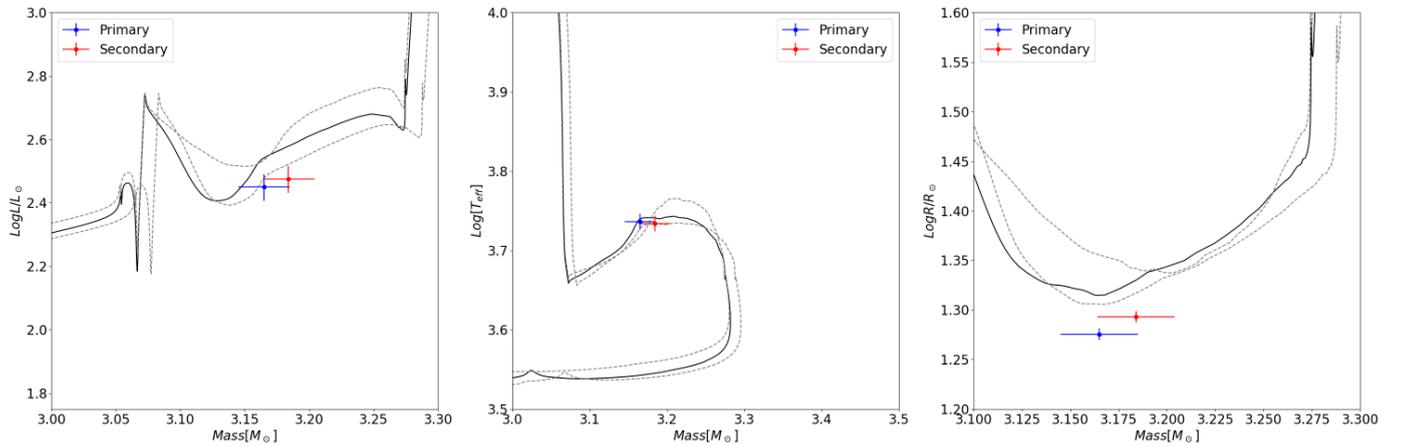

**Fig. B.3.** Same as Figure B.1 but for the system LMC554.